\documentclass[a4paper]{jpconf}
\usepackage{graphicx}
\usepackage{wrapfig}
\usepackage{amssymb}
\usepackage{amsmath}

\begin{document}
\title{Big-bang nucleosynthesis with a long-lived
CHAMP including He4 spallation
process}

\author{Masato Yamanaka$^1$, Toshifumi Jittoh$^2$, Kazunori Kohri$^1$, 
Masafumi Koike$^2$, Joe Sato$^2$, Kenichi Sugai$^2$, and Koichi  Yazaki$^{3, 4}$}

\address{$^1$Theory Center, Institute of Particle and Nuclear Studies,
KEK (High Energy Accelerator Research Organization),
1-1 Oho, Tsukuba 305-0801, Japan}
\address{$^2$Department of Physics, Saitama University, 
Shimo-okubo, Sakura-ku, Saitama, 338-8570, Japan}
\address{$^3$Hashimoto Mathematical Physics Laboratory , 
Nishina Accelerator Research Center, 
RIKEN, Wako, Saitama 351-0198, Japan}
\address{$^4$Yukawa Institute for Theoretical Physics, 
Kyoto University, Kyoto 606-8502, Japan}

\ead{masato.yamanaka@kek.jp}

\begin{abstract}
We propose helium-4 spallation processes induced by long-lived 
stau in supersymmetric standard models, and investigate an 
impact of the processes on light elements abundances.  
We show that, as long as the phase space of helium-4 spallation 
processes is open, they are more important than stau-catalyzed 
fusion and hence constrain the stau property.  
This talk is based on works~\cite{Jittoh:2011ni}. 
\end{abstract}

\section{Introduction} 

Long-lived charged massive particles (CHAMPs) will play interesting roles in 
the Big-Bang Nucleosynthesis (BBN). The light nuclei will interact not only with 
the CHAMPs during the BBN processes~\cite{Pospelov:2006sc, 
Jittoh:2007fr, Bird:2007ge, Jittoh:2008eq, Kamimura:2008fx, Jittoh:2010wh}, 
but also with the decay products of the CHAMPs in the post-BBN 
era~\cite{Kawasaki:2004yh, Kawasaki:2004qu}.
The standard BBN will thus be altered, and so is the abundance
of the light elements at the present time. One can thus constrain the models 
beyond the Standard Model by evaluating their prediction on the light elements 
abundance and comparing it with the current observations. We can then give 
stringent predictions for the forthcoming experiments and observations according 
to these constraints.

The Standard Model extended with supersymmetry (SUSY) is one of the models 
that can accommodate such long-lived CHAMPs.
With the $R$-parity conservation, the lightest SUSY particle
(LSP) is stable and become a cold dark matter.
Interestingly, it can offer a long-lived CHAMPs if the LSP is the bino-like
neutralino $\tilde{\chi}_{1}^{0}$.
Coannihilation mechanism is required to account for the dark matter abundance in
this case~\cite{Griest:1990kh}, where the LSP and the next-lightest SUSY
particle (NLSP) are almost degenerate in mass.
Staus, denoted by $\tilde{\tau}$ and a possible candidate of the NLSP, can 
acquire a long lifetime when the mass difference with the LSP is less than the 
mass of tau leptons. This is due to the phase space suppression of the final 
state that necessarily consists of three particles or more.
Noting that such long-lived staus will be copious during the
BBN~\cite{Profumo:2004qt,Jittoh:2005pq}, we have shown 
in~\cite{Jittoh:2007fr, Jittoh:2008eq, Jittoh:2010wh} 
that their presence indeed alters the prediction of the standard BBN and
possibly solve the discrepancy of the lithium abundance in the
Universe through the internal conversion reactions.

In this talk, we propose new reactions
\begin{subequations}
\begin{align}
    (\tilde \tau \hspace{0.2mm} ^4\text{He}) 
    &\to \tilde \chi_1^0 + \nu_\tau + \text{t} + \text{n},
    \label{eq:spal-tn}
    \\
    (\tilde \tau \hspace{0.2mm} ^4\text{He}) 
    &\to \tilde \chi_1^0 + \nu_\tau + \text{d} + \text{n} + \text{n}, 
    \label{eq:spal-dnn}
    \\ 
    (\tilde \tau \hspace{0.2mm} ^4\text{He}) 
    &\to \tilde \chi_1^0 + \nu_\tau + \text{p} + \text{n} + \text{n} + \text{n},
    \label{eq:spal-pnnn}
\end{align}
    \label{spa_1}
\end{subequations}
in which $(\tilde \tau \, ^4\text{He})$ represents a bound state of a stau and
$\mathrm{^{4}He}$ nucleus.
Reaction (\ref{spa_1}) is essentially a spallation of the $\mathrm{^{4}He}$
nucleus, producing a triton t, a deuteron d, and neutrons n.
Presence of such spallation processes has been ignored so far
due to the na\"{i}ve expectation that
the rate of the stau-catalyzed fusion~\cite{Pospelov:2006sc}
\begin{equation}
  (\tilde\tau \, \mathrm{^{4}He}) + \mathrm{d}
  \to
  \tilde \tau + \mathrm{^{6}Li}
  \label{eq:Pospelov}
\end{equation}
is larger than the reaction (\ref{spa_1}).
Indeed, the cross section of Eq.~(\ref{eq:Pospelov}) is much larger than that
of $^4$He + d $\to$ $^6$Li + $\gamma$ by (6 -- 7) orders of
magnitude~\cite{Hamaguchi:2007mp}.

We point out that this expectation is indeed na\"{i}ve; the reaction 
Eq.~({\ref{spa_1}}) is more effective than Eq.~({\ref{eq:Pospelov}) 
as long as the spallation processes are kinematically allowed. 
The former reaction rapidly occurs due to the large overlap of their wave 
functions in a bound state.
On the other hand, the latter proceeds slowly since it requires an 
external deuteron which is sparse at the BBN era.  The overproduction of 
t and d is more problematic than that of $\mathrm{^{6}Li}$.
This puts new constraints on the parameters of the minimal supersymmetric 
standard model (MSSM). Note that there 
is no reaction corresponding to Eq.~({\ref{spa_1}}) in the gravitino LSP 
scenario~\cite{Feng:2003xh, Buchmuller:2007ui}.

\section{Spallation of helium 4}  \label{sec:int} 

\begin{wrapfigure}{r}{60mm}
\begin{center}
  \includegraphics[width=50mm]{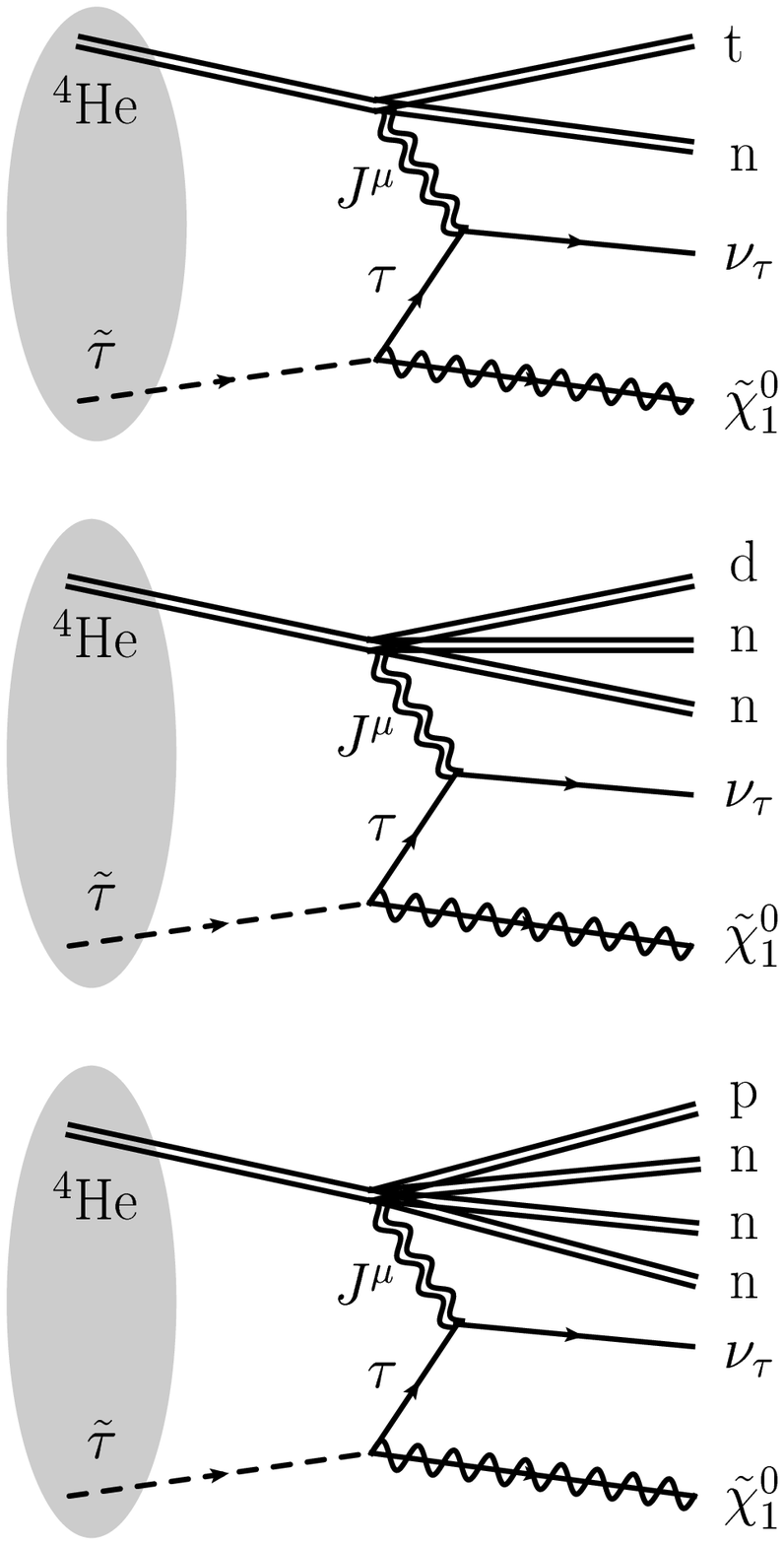}
  \caption{$^4$He spallation processes.}
\end{center}
\end{wrapfigure}

The $^4$He spallation processes of Eq.~(\ref{spa_1}) is described by the Lagrangian
\begin{equation}
\begin{split}
    \mathcal{L}
    = \tilde \tau^* \overline{ \tilde \chi_1^0 } 
    (g_\text{L} P_\text{L} + g_\text{R} P_\text{R}) \tau   
    + \sqrt{2} G_{\textrm{F}}
    \nu_\tau \gamma^\mu P_\text{L} \tau J_\mu    
    + \text{h.c.}, 
\end{split}     \label{Lag}
\end{equation} 
where $G_{\textrm{F}} = 1.166 \times 10^{-5} \mathrm{GeV^{-2}}$ is the 
Fermi coupling constant, $P_{\text{L(R)}}$ represents the chiral projection operator, 
and $J_{\mu}$ is the weak current. The effective coupling constants $g_\text{L}$ 
and $g_\text{R}$ are given by 
\begin{equation}
\begin{split}
    g_\text{L} 
    = \frac{g}{\sqrt{2} \cos\theta_\text{W}}  
    \sin\theta_\text{W} \cos\theta_\tau ,   ~ 
    g_\text{R} 
    = \frac{\sqrt{2} g}{\cos\theta_\text{W}}  
    \sin\theta_\text{W} \sin\theta_\tau 
    \mathrm{e}^{i \gamma_\tau}, 
\end{split}
\end{equation}
where $g$ is the $SU(2)_\text{L}$ gauge coupling constant and
$\theta_{\textrm{W}}$ is the Weinberg angle. The mass eigenstate of staus 
is given by the linear combination of $\tilde \tau_{\textrm{L}}$ and $\tilde 
\tau_{\textrm{R}}$, the superpartners of left-handed and right-handed tau 
leptons, as
\begin{equation}
\begin{split}
    \tilde \tau 
    = \cos\theta_\tau \tilde \tau_\text{L} 
    + \sin\theta_\tau \mathrm{e}^{-i \gamma_\tau} 
    \tilde \tau_\text{R} . 
\end{split}
\end{equation}
Here $\theta_\tau$ is the left-right mixing angle of staus and $\gamma_\tau$ 
is the CP violating phase.

\clearpage
\subsection{$(\tilde \tau \hspace{0.5mm} ^4\text{He}) \to 
\tilde \chi_1^0 + \nu_\tau + \text{t} + \text{n}$ } \label{sec:Tn} 

First we consider the process of Eq.~({\ref{eq:spal-tn}}).
The rate of this process is expressed as 
\begin{equation}
\begin{split}
   \frac{1}{\tau_\text{tn}} 
   = \frac{1}{|\psi|^2 \cdot 
   \sigma v_{\textrm{tn}}}, 
\end{split}     \label{time_Tn}
\end{equation}
where $|\psi|^2$ stands for the overlap of the wave functions of the stau
and the $\mathrm{^4He}$ nucleus.
We estimate the overlap by
\begin{equation}
\begin{split}
   |\psi|^2 = \frac{(Z \alpha m_\text{He})^3}{\pi},
\end{split}     \label{overlap}
\end{equation}
where $Z$ and $m_\text{He}$ represent the atomic number and the mass of $^4$He,
respectively, and $\alpha$ is the fine structure constant.
We assumed that the stau is pointlike particle and is much heavier than
$\mathrm{^{4}He}$ nucleus so that the reduced mass of the bound
state is equal to the mass of $\mathrm{^4He}$ nucleus itself.
The cross section of the elementary process for this reaction is denoted by
$\sigma v_\text{tn}$ and calculated as
\begin{equation}
\begin{split}
   \sigma v_\text{tn}  
   &\equiv 
   \sigma v \bigl( (\tilde \tau ^4\text{He}) \to 
   \tilde \chi_1^0 \nu_\tau \text{tn} \bigr)
   \\ &
   = 
   \frac{1}{2E_{\tilde \tau}} \int
   \frac{d^3 \boldsymbol{p}_\nu}{(2\pi)^3 2E_\nu} 
   \frac{d^3 \boldsymbol{p}_{\tilde \chi}}{(2\pi)^3 2E_{\tilde \chi}} 
   \frac{d^3 \boldsymbol{q}_\text{n}}{(2\pi)^3}
   \frac{d^3 \boldsymbol{q}_\text{t}}{(2\pi)^3} 
   \\ & \hspace{3em} \times 
   \bigl| \mathcal{M} \bigl( (\tilde \tau ^4\text{He}) \to 
   \tilde \chi_1^0 \nu_\tau \text{tn} \bigr) \bigr|^2 
   \\ & \hspace{3em} \times 
   (2 \pi)^4 \delta^{(4)} 
   (p_{\tilde \tau} + p_\text{He} - p_\nu - q_\text{t} - q_\text{n}). 
\end{split}     \label{cross_tn_1}
\end{equation}
Here $p_i$ and $E_i$ are the momentum and the energy of the particle 
species $i$, respectively.

We briefly show the calculation of the amplitude of this process.
The amplitude is deconstructed as
\begin{equation}
\begin{split}
   \mathcal{M} \bigl( (\tilde \tau ^4\text{He}) \to 
   \tilde \chi_1^0 \nu_\tau \text{tn} \bigr)
   &= \langle \text{t \hspace{-1.9mm} n} \hspace{0.5mm} 
   \tilde \chi_1^0 \hspace{0.5mm} \nu_\tau 
   |\mathcal{L}_{\textrm{int}}| 
   ^4\text{He} \hspace{0.5mm} \tilde \tau \rangle 
   \\ & = 
   \langle \text{t \hspace{-1.9mm} n} 
   |J^\mu| 
   ^4\text{He} \rangle  ~ 
   \langle \tilde \chi_1^0 \hspace{0.5mm} \nu_\tau 
   |j_\mu| 
   \tilde \tau \rangle.
   \label{ampli_Tn}
\end{split}
\end{equation}
The weak current $J_{\mu}$ consists of a vector current $V_{\mu}$ and an
axial vector current $A_{\mu}$ as $J_{\mu} = V_{\mu} + g_{\textrm{A}} 
A_{\mu}$, where $g_{\textrm{A}}$ is the axial coupling constant.
The relevant components of the currents in this reaction are $V^{0}$ 
and $A^{i}$ ($i = 1, 2, 3$).
We take these operators as a sum of a single-nucleon  operators as
\begin{equation}
  V^{0}
  =
  \sum_{a = 1}^{4} \tau_{a}^{-}
  \mathrm{e}^{\mathrm{i}\boldsymbol{q} \cdot \boldsymbol{r}_{a}}
  \, , \quad
  A^{i}
  =
  \sum_{a = 1}^{4} \tau_{a}^{-} \sigma_{a}^{i}
  \mathrm{e}^{\mathrm{i}\boldsymbol{q} \cdot \boldsymbol{r}_{a}}
  \, ,
\end{equation}
where $\boldsymbol{q}$ is the momentum carried by the current, 
$\boldsymbol{r}_{a}$ is the spatial coordinate of the $a$-th nucleon ($a \in 
\{1, 2, 3, 4 \}$), and $\tau_a^-$ and $\sigma_{a}^{i}$ denote the isospin 
ladder operator and the spin operator of the $a$-th nucleon, respectively. 
Each component leads to a part of hadronic matrix element:
\begin{equation}
\begin{split}
   \langle \text{tn} | V^0 | ^4\text{He} \rangle 
   & = \sqrt{2} \mathcal{M}_{\textrm{tn}} , 
   \\ 
   \langle \text{tn} | g_{\textrm{A}} A^+ | ^4\text{He} \rangle 
   & = \sqrt{2} g_{\textrm{A}} \mathcal{M}_{\textrm{tn}} , 
   \\ 
   \langle \text{tn} | g_{\textrm{A}} A^- | ^4\text{He} \rangle 
   & = - \sqrt{2} g_{\textrm{A}} \mathcal{M}_{\textrm{tn}}, 
   \\
   \langle \text{tn} | g_{\textrm{A}} A^3 | ^4\text{He} \rangle 
   & = - \sqrt{2} g_{\textrm{A}} \mathcal{M}_{\textrm{tn}},
   \label{had_mat} 
\end{split}
\end{equation}
where $A^\pm = (A^1 \pm i A^2)/\sqrt{2}$.
Given the relevant wave functions of a $\mathrm{^4He}$ nucleus, a triton, and a
neutron (see Appendix in Ref.~\cite{Jittoh:2011ni}), we obtain the hadronic matrix 
element as
\begin{equation}
\begin{split}
   &
   \mathcal{M}_{\textrm{tn}} 
   = 
   \biggl(
   \frac{128 \pi}{3} 
   \frac{a_\text{He} a_\text{t}^2}{(a_\text{He} + a_\text{t})^4}
   \biggr)^{3/4} 
   \biggl\{ 
   \exp \biggl[ - \frac{\boldsymbol{q}_\text{t}^2}{3 a_\text{He}} \biggr] 
   - \exp \biggl[ - \frac{\boldsymbol{q}_\text{n}^2}{3 a_\text{He}} 
   - \frac{(\boldsymbol{q}_\text{t} + \boldsymbol{q}_\text{n})^2}
   {6 (a_\text{He} + a_\text{t})} \biggr]
   \biggr\}.
   \label{amp_part} 
\end{split}
\end{equation}
Here $\boldsymbol{q}_\text{t}$ and $\boldsymbol{q}_\text{n}$ are three-momenta
of the triton and the neutron, respectively, and $a_\text{He}$ and $a_\text{t}$
are related to the mean square matter radius $R_\text{m}$ by
\begin{equation}
\begin{split}
   a_\text{He} 
   = \frac{9}{16} \frac{1}{ (R _{\text{m}})  _{\text{He}} ^2}, ~
   a_\text{t} 
   = \frac{1}{2} \frac{1}{ (R _{\text{m}}) _{ \text{t}}  ^2}
   \label{a_r} .   
\end{split}
\end{equation}
We list in Table \ref{table:input} input values of the matter radius for the numerical calculation 
in this article.

\begin{table}
\caption{Input values of the matter radius $R_\text{mat}$ 
for d, t, and $^4\text{He}$, the magnetic radius 
$R_\text{mag}$ for p and n, nucleus mass $m_X$, excess 
energy $\Delta_X$ for the nucleus $X$, and each reference. 
\label{table:input}}
\begin{center}
\begin{tabular}{llll} 
\br
nucleus  
& $R_\text{mat(mag)}$ [fm]/[GeV$^{-1}$] ~
& $m_X$ [GeV] 
& 
$\Delta_X$ [GeV] 
\\[0.7mm] \hline
p
& 0.876 / 4.439 \cite{Borisyuk:2009mg}
& 0.9383 \ \cite{P.Mohr}
& $6.778 \times 10^{-3}$ \ \cite{TOI}
\\[0.7mm] \hline
n
& 0.873 / 4.424 \cite{Kubon:2001rj}
& 0.9396 \ \cite{P.Mohr}
& $8.071 \times 10^{-3}$ \ \cite{TOI}
\\[0.7mm] \hline
d          
& 1.966 / 9.962 \ \cite{Wong:1994sy}
& 1.876 \ \cite{TOI}  ~
& $1.314 \times 10 ^{-2}$ \ \cite{TOI}
\\[0.7mm] \hline
t           
& 1.928 / 9.770 \ \cite{Yoshitake}  
& 2.809 \ \cite{TOI}  
& $1.495 \times 10 ^{-2}$ \ \cite{TOI}
\\[0.7mm] \hline
$^4$He 
& 1.49 / 7.55 \ \cite{Egelhof2001307}
& 3.728  \ \cite{TOI}
& $2.425 \times 10 ^{-3}$ \ \cite{TOI}
\\[0mm] \br
\end{tabular}
\end{center}
\end{table}

The remaining part is straightforwardly calculated to be
\begin{equation}
\begin{split}
   &
   | \langle \tilde \chi_1^0 \hspace{0.5mm} \nu_\tau 
   |j_0| \tilde \tau \rangle |^2
   = 
   | \langle \tilde \chi_1^0 \hspace{0.5mm} \nu_\tau 
   |j_z| \tilde \tau \rangle |^2 
   = 
   4 G_\text{F}^2 |g_\text{R}|^2 
   \frac{m_{\tilde \chi_1^0} E_\nu}{m_\tau^2},
   \\ &
   | \langle \tilde \chi_1^0 \hspace{0.5mm} \nu_\tau 
   |j_\pm| \tilde \tau \rangle |^2 
   = 4 G_\text{F}^2 |g_\text{R}|^2 
   \frac{m_{\tilde \chi_1^0} E_\nu}{m_\tau^2} 
   \biggl( 
   1 \mp \frac{p_\nu^z}{E_\nu}
   \biggr) , 
   \label{leptonic_part}
\end{split} 
\end{equation}
where $E_\nu$ and $p_\nu^z$ are the energy and the $z$-component of the 
momentum of the tau neutrino, respectively. 
We assumed that the stau and the neutralino are non-relativistic.
This equation includes not only all the couplings such as $G_{\textrm{F}}$,
$g_{\textrm{L}}$, and $g_{\textrm{R}}$, but also the effect of the virtual tau
propagation in the Fig.~1.
Note here that $g_{\textrm{L}}$ coupling does not contribute.
This is because the virtual tau ought to be left-handed at the weak current,
and it flips its chirality during the propagation since the transferred
momentum is much less than its mass.

Combining hadronic part with the other part, we obtain the squared amplitude as
\begin{equation}
\begin{split}
   \bigl| \mathcal{M} \bigl( (\tilde \tau ^4\text{He}) \to 
   \tilde \chi_1^0 \nu_\tau \text{tn} \bigr) \bigr|^2 
   = \frac{8 m_{\tilde \chi_1 ^0} G_{\text{F}}^2 |g_{\text{R}}|^2}{m_{\tau}^2}
   (1 + 3 g_{A}^2) E_{\nu} 
   |\mathcal{M}_{\textrm{tn}}|^2  . 
   \label{squ_amp}
\end{split} 
\end{equation}
Integrating on the phase space of the final states, we obtain the cross section as  
\begin{equation}
\begin{split}
   \sigma v_\text{tn} 
   = 
   \frac{8}{\pi^2} \biggl( \frac{32}{3 \pi} \biggr)^{3/2}
   g^2 \tan^2\theta_W \sin^2\theta_\tau (1+3g_A^2) G_F^2 
   \Delta_\text{tn}^4 
   \ \frac{m_\text{t} m_\text{n}}{m_{\tilde \tau} m_\tau^2} \ 
   \frac{a_\text{He}^{3/2} a_\text{t}^3}{(a_\text{He} + a_\text{t})^5} 
   ~I_\text{tn} , 
   \label{cross_tn_2} 
\end{split} 
\end{equation}
Here $I_\text{tn}$ is the numerically calculated factor including the information 
of phase space for this process. Analytic form of $I_\text{tn}$ is shown in 
Ref.~\cite{Jittoh:2011ni}, and numerical result is depicted in Fig.~2. 
$\Delta_\text{tn}$, $k_\text{t}$, and $k_\text{n}$ are defined as 
\begin{equation}
\begin{split}
   &
   \Delta_\text{tn} \equiv \delta m + \Delta_\text{He} - \Delta_\text{t} 
   - \Delta_\text{n} - E_\text{b} ,
   \\ &
   k_\text{t} \equiv \sqrt{2 m _{\text{t}} \Delta_\text{tn} } \ ,~~~  
   k_\text{n} \equiv \sqrt{2 m _{\text{n}} \Delta_\text{tn} } \ ,
   \label{k_tn} 
\end{split} 
\end{equation}
where $\Delta_X$ is the excess energy of the nucleus $X$, and $E_\text{b}$ 
is the binding energy of $(\tilde \tau \hspace{0.3mm} ^4\text{He})$ system.

\section{Comparing the rate of spallation reaction with that of stau-catalyzed fusion}   
\label{sec:rate}   

\begin{figure}[t!]
\begin{minipage}{18.5pc}
\includegraphics[width=18.5pc]{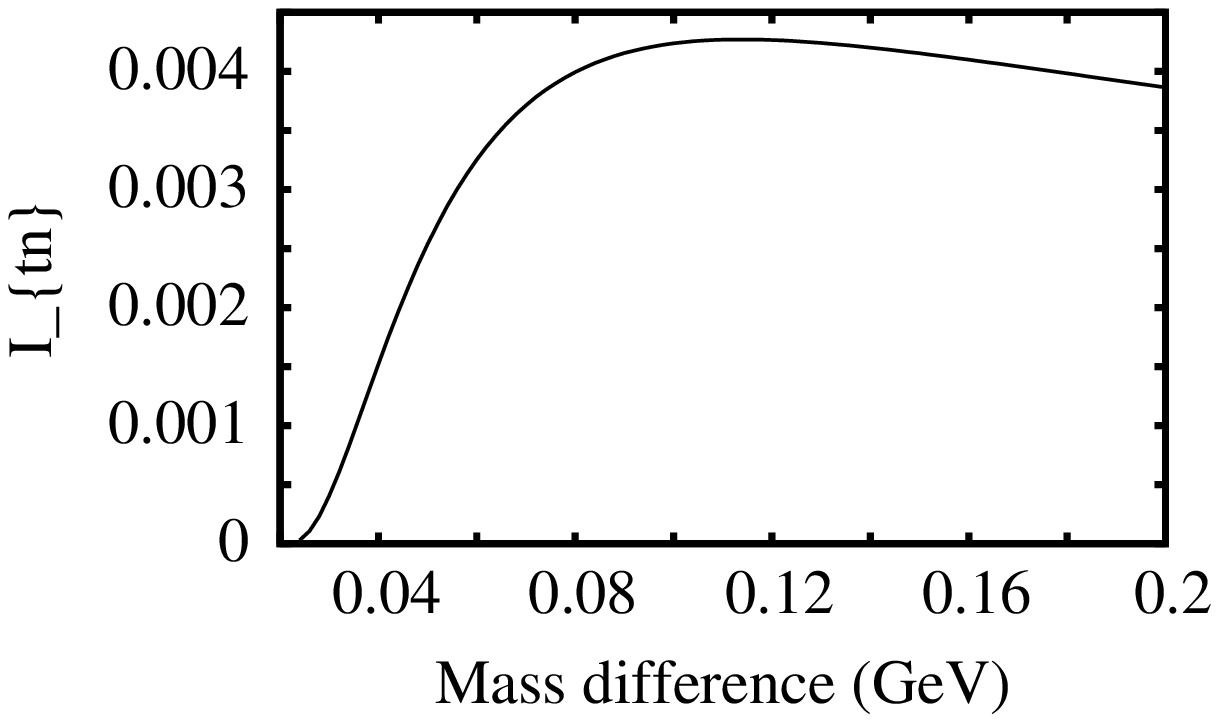}
\caption{\label{Itn}Factor $I_\text{tn}$ in Eq.~(\ref{cross_tn_2}) as a 
function of mass difference between the stau and the neutralino. 
Here we took $m_{\tilde \tau} = 350$GeV, $\sin\theta_\tau 
= 0.8$, and $\gamma_\tau = 0$.}
\end{minipage}\hspace{1pc}%
\begin{minipage}{18.5pc}
\includegraphics[width=18.5pc]{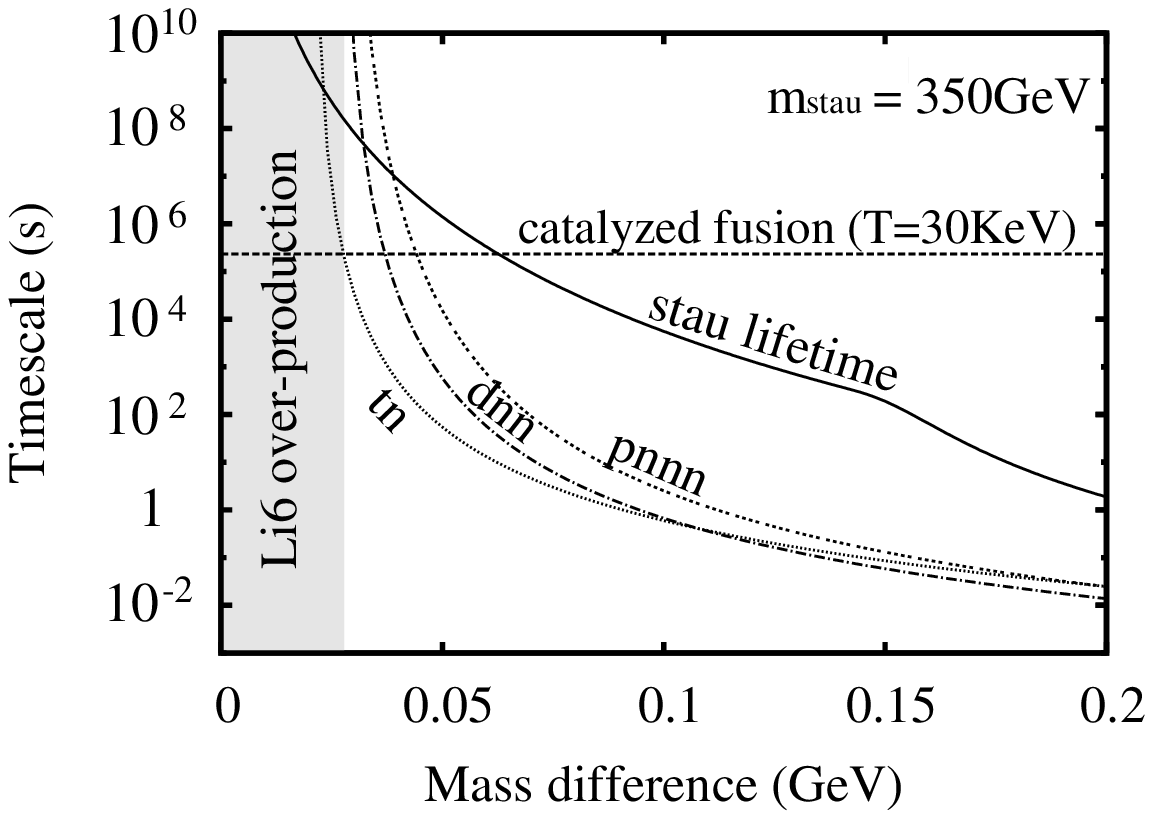}
\caption{\label{fig:time}Timescale of spallation processes as a function 
of $\delta m$ and the stau-catalyzed fusion at the 
universe temperature $T=30$keV~\cite{Hamaguchi:2007mp}. 
The lifetime of free $\tilde \tau$ (solid line) is also depicted.  
Here we took $m_{\tilde \tau} = 350$GeV, $\sin\theta_\tau 
= 0.8$, and $\gamma_\tau = 0$.}
\end{minipage} 
\end{figure}

We compare the rate of the spallation and that of the stau-catalyzed fusion. 
We first note that the rate of stau-catalyzed fusion strongly depends on the
temperature~\cite{Hamaguchi:2007mp}, and we fix the reference 
temperature to be $30\mathrm{keV}$.
Staus begin to form a bound state with $^4$He at this temperature, which 
corresponds to cosmic time of $10^{3}\mathrm{s}$. Thus the bound 
state is formed when the lifetime of staus is longer than $10^{3}\mathrm{s}$.

Figure~\ref{fig:time} shows the timescale of the spallation processes as a 
function of $\delta m$. The lifetime of free stau is plotted by a solid line. 
We took the reference values of $m_{\tilde \tau} = 350$GeV, 
$\sin\theta_\tau = 0.8$, and $\gamma_\tau = 0$. The inverted rate of 
the stau-catalyzed fusion at the temperature of $30\mathrm{keV}$ is also 
shown by the horizontal dashed line.
Once a bound state is formed, as long as the phase space of spallation 
processes are open sufficiently that is $\delta m \gtrsim 0.026$GeV, 
those processes dominate over other processes.  There $\tilde \tau$ 
property is constrained to evade the over-production of d and/or t. 
For $\delta m \lesssim 0.026$GeV, the dominant process of $(\tilde \tau 
\hspace{0.3mm} ^4\text{He})$ is stau-catalyzed fusion, since the 
free $\tilde \tau$ lifetime is longer than the timescale of stau-catalyzed 
fusion. Thus light gray region is forbidden due to the over-production 
of $^6$Li.

This interpretation of Fig.2 is not much altered by varying the parameters 
relevant with $\tilde \tau$.  First cross sections of spallation processes are 
inversely proportional to $m_{\tilde \tau}$,  and then the timescale of each 
process linearly increases as $m_{\tilde \tau}$ increases. Thus, even when 
$m_{\tilde \tau}$ is larger than $m_{\tilde \tau} = 350$GeV by up to a 
factor of ten, the region of $^6$Li over-production scarcely changes.  
Next we point out that our result depend only mildly on the left-right mixing 
of the stau.  Indeed, cross section of the $^4 \text{He}$ spallation is 
proportional to $\sin ^2 \theta _{\tau}$.  Its order of magnitude will not 
change as long as the right-handed component is significant.

\section{Summary}  \label{sec:sum} 

Long-lived charged massive particles provides some exotic nuclear 
reactions in the big bang nucleosynthesis. So it is important for 
understanding the property of long-lived charged massive particles 
to understand what type of exotic nuclear induce over-production 
(-destruction). 
Newly included in the present work is the spallation of the $^4$He 
in the stau-$^4$He bound state given in Eq.(1). 
This process is only present in the model which predicts the long-lived 
charged particles due to the phase space suppression with 
the weakly interacting daughter particle.

We calculated the rate of the helium-4 spallation processes 
analytically, and compared it with that of catalyzed fusion.   
We found that the spallation of $^4$He nuclei dominate over the 
catalyzed fusion as long as the phase space of the spallation 
processes are open and hence the property of long lived stau 
is constrained from avoiding the overproduction of a deuteron 
and/or a triton.

\section*{Acknowledgments}   

This  work was supported in part by the Grant-in-Aid for the Ministry
of Education, Culture, Sports, Science, and Technology, Government  of
Japan, No.~21111006, No.~22244030, No.~23540327(K.K.), 
No. 24340044 (J.S.),  and No. 23740208 (M.Y.).


\end{document}